\documentclass[fleqn,usenatbib]{mnras}

\usepackage{newtxtext,newtxmath}

\usepackage[T1]{fontenc}

\DeclareRobustCommand{\VAN}[3]{#2}
\let\VANthebibliography\thebibliography
\def\thebibliography{\DeclareRobustCommand{\VAN}[3]{##3}\VANthebibliography}

\usepackage{graphicx}	
\usepackage{amsmath}	

\title[Fuzzy dark matter and filament profiles]{Searching for signatures of fuzzy dark matter in cosmic filament profiles}

\author[C. J. O'Kane et al.]{
Callum J. O'Kane$^{1,2}$\thanks{E-mail: callum.o'kane@nottingham.ac.uk},
Alfonso Arag\'{o}n-Salamanca$^{1}$,
Ulrike Kuchner$^{1}$,
and Meghan E. Gray$^{1}$
\\
$^{1}$School of Physics and Astronomy, University of Nottingham, Nottingham NG7 2RD, UK\\
$^{2}$Isaac Newton Group of Telescopes, Apartado 321, E-38700 Santa Cruz de la Palma, Spain
}

\date{Accepted XXX. Received YYY; in original form ZZZ}

\pubyear{2024}

\begin{document}
\label{firstpage}
\pagerange{\pageref{firstpage}--\pageref{lastpage}}
\maketitle

\begin{abstract}
Current observations reveal persistent tensions with the standard cold dark matter paradigm, raising the question of whether these can be explained by baryonic physics alone or require alternative dark matter models. One such alternative is fuzzy (or wave) dark matter, consisting of ultralight particles with mass $m \sim 10^{-22}$ eV and de Broglie wavelengths on kpc to Mpc scales, which may give rise to large--scale interference patterns in non-linear structures around the cosmic web, such as filaments and clusters. In this work, we search for possible signatures of these interference fringes by investigating periodicities in the distribution of galaxies around cosmic web filaments. To demonstrate our methodology, we compare the filament profiles to a simple model that includes a periodic component of the form $A\cos(2\pi d/ \lambda)$, where $A$ is the maximum density contrast (amplitude) of the periodic component, with wavelength $\lambda=\lambda_0 \cos\theta$ for some face-on wavelength $\lambda_0$ inclined at an angle $\theta$ to the line of sight. Exploiting the large Sloan Digital Sky Survey (SDSS) Main Galaxy Sample, we analyse a sample of 4,394 filaments from the Tempel et al. filament catalogue, each containing at least 10 member galaxies. We find a vast portion of the parameter space is consistent with the observations at the $2\sigma$ level, including all models with $A = 0$ (no periodicity). We identify a region of the parameter space in tension with the observations, allowing us to exclude values of $A > 0.16 \lambda_0 + 0.18$ for $0.2\,\text{Mpc}\,\lesssim \lambda_0 \lesssim 2\,\text{Mpc}$ at the $3\sigma$ level, demonstrating the ability to test models of filament dark matter structure using this methodology.

\end{abstract}




\section{Introduction}
Despite the considerable success of the cold dark matter (CDM) cosmological paradigm, its predictions remain in tension with a number of observations \citep{weinberg_cold_2015, bullock_small-scale_2017}. These discrepancies predominantly arise on small scales ($\lesssim1\,\text{Mpc}$, $\lesssim10^{11}\,\text{M}_{\sun}$)\footnote{For a comprehensive review of these small-scale issues, see \cite{bullock_small-scale_2017} and references within.}, with some of the most noteable examples being the ``cuspy core'', ``missing satellite'' and ``Too-Big-to-Fail'' problems \citep{moore_evidence_1994,klypin_where_1999,boylan-kolchin_too_2011,boylan-kolchin_milky_2012}. Whilst substantial effort has been devoted to reconciling these tensions through baryonic physics (e.g. \citealt{navarro_cores_1996,van_den_bosch_constraints_2000,ferrero_dark_2012,governato_cuspy_2012,ostriker_mind_2019}), they have also motivated the exploration of alternative dark matter models.

One particularly well-studied alternative is fuzzy (or wave) dark matter, in which the dark matter consists of ultralight particles ($m \sim 10^{-22}$ eV) with sufficiently high number density to best be described as a classical wave (e.g. \citealt{hu_cold_2000, hui_wave_2021}). This stands in contrast to the CDM framework, where dark matter is modelled as a collection of collisionless particles.

Ultralight dark matter has received increasing attention as a viable alternative to CDM, with a growing body of work in both astrophysics and particle physics exploring both its properties and its phenomenology. In this work, we focus on the model proposed in the work of \cite{hu_cold_2000}, which considers ultralight particles with masses of $\sim 10^{-22}$ eV, named Fuzzy dark matter (FDM hereafter). Such masses introduce a characteristic length scale, below which gravitational collapse is suppressed by quantum pressure arising from the uncertainty principle \citep{hu_cold_2000,kousha_effective_2023}. For particle masses of $m \sim 10^{-22}$ eV, the corresponding de Broglie wavelength is of the order of kpc -- Mpc, and may therefore produce potentially observable effects on observable scales (e.g. \citealt{boldrini_-situ_2025, szpilfidel_fuzzy_2025, ellis_constraints_2025,may_updated_2025, liu_constraining_2025, wang_fuzzy_2025, koo_head-collisions_2025, palencia_signatures_2025}). The introduction of a minimum length scale in FDM offers a possible resolution to several small-scale challenges faced by CDM, motivating a growing effort to both simulate FDM cosmologies and search for their observational signatures.

Numerical simulations of FDM are computationally demanding, especially in comparison to CDM simulations, as they must resolve both small spatial and temporal scales in order to accurately capture wave-like effects \citep{schive_fuzzy_2025}. Consequently, existing simulations are typically limited to relatively small volumes and short cosmic timescales (e.g. \citealt{schive_cosmic_2014,mocz_first_2019,li_numerical_2019,mocz_galaxy_2020,may_structure_2021, kulkarni_if_2022}). Despite these limitations, such simulations consistently predict striking departures from their CDM counterparts. One notable prediction is the presence of interference fringes in the distribution of dark matter. These fringes are a product of constructive and destructive interference associated with the large de Broglie wavelength of FDM particles. One regime in which these features are prevalent is the non-linear regime of the cosmic web, within filaments and clusters. Such features are clearly visible in the dark matter distributions of simulations (e.g. Figure 1 of \citealt{may_structure_2021}, Figure 1 of \citealt{mocz_first_2019} and Figure 1 of \citealt{schive_cosmic_2014}). The presence of interference patterns is a distinctive prediction of wave-like dark matter models and would constitute a compelling observational signature if detected \citep{schive_cosmic_2014, mocz_first_2019,hui_wave_2021,ferreira_ultra-light_2021}.

Searches for observational signatures of FDM have progressively constrained the allowed particle mass range, employing a variety of complementary approaches. For example, studies of stellar dynamics in Milky Way dwarf spheroidal galaxies place a lower bound of $0.4 \times 10^{-22}$ eV \citep{gonzalez-morales_unbiased_2017}. Analyses of the one-dimensional 
Lyman-$\alpha$ forest flux power spectrum at high redshift yield more stringent limits, for example the  $m > 2 \times 10^{-20}$ eV limit reported in \cite{rogers_strong_2021}. More recently, \cite{liu_joint_2025} have reported a similar lower bound of $m > 1.75 \times 10^{-20}$ eV based on the satellite populations of the Milky Way and Andromeda.
\begin{figure}
	\includegraphics[width=\columnwidth,keepaspectratio]{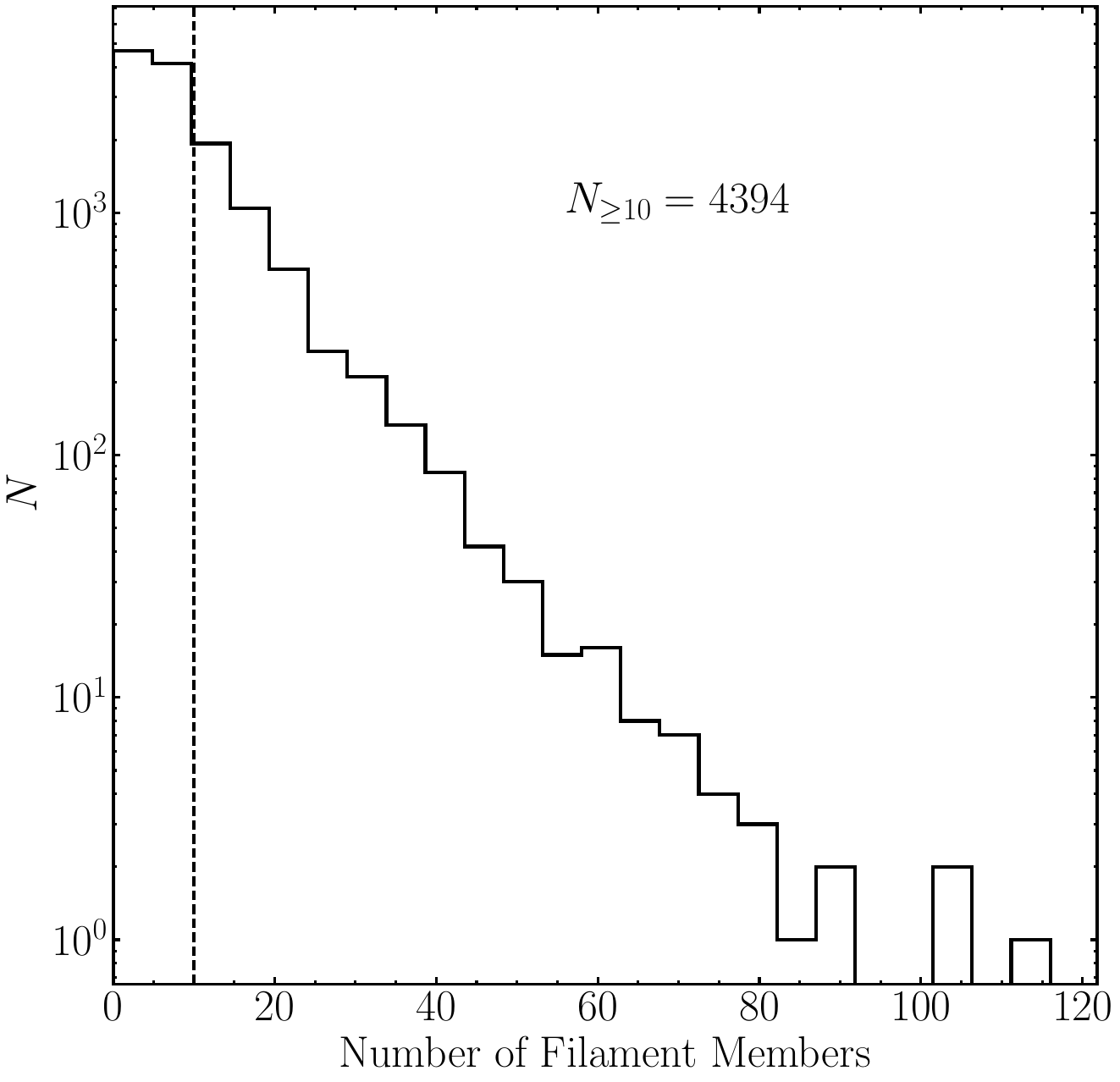}
    \caption{The number of filaments as a function of galaxy members. Galaxy members are defined as galaxies within 1.5\,Mpc of a given filament spine which do not reside within close proximity to a group/cluster (distances larger than $R_{180, \text{group}}$ and $2.5R_{180,\text{cluster}} $; see \autoref{Group and Cluster Catalogue}). The vertical dashed line denotes the 10 member limit; only filaments with 10 or more members are considered in this work (see \autoref{Sec: Observational Power Spectra}), resulting in a filament sample of size 4,394. }
    \label{fig:FilamentMembers}
\end{figure}
Whilst these constraints appear to disfavour the parameter space in which FDM is expected to have prominent astrophysical effects, it is important to note that the interplay between FDM and baryonic physics remains poorly understood, owing to the substantial challenges involved in simulating both simultaneously. Moreover, work such as that of \cite{koo_head-collisions_2025} find that, through simulations of head-on collisions of FDM subhalos, the kinetic energy dissipation in an FDM model is much different from that of CDM, and could perhaps provide an explanation for some colliding systems such as Abell 520, which CDM cannot completely explain \citep{mahdavi_dark_2007}. Furthermore, the recent work of \cite{hou_flux-ratio_2026} investigated the flux-ratios of 17 cusp quasars, finding that the implied dark matter substructure is better explained with an FDM model than CDM. Overall, the current observational and theoretical landscape does not yet allow FDM to be conclusively ruled out or be confirmed.

Despite being a distinctive and potentially observable feature of fuzzy dark matter, the search for interference fringes in cosmic web filaments has, to the best of our knowledge, remained unexplored. This is perhaps unsurprising, as the density contrast in the cosmic web is strongest at late cosmic times, whereas current FDM simulations which incorporate wave-like behaviour are typically limited to early epochs. For example, the large scale simulations of \citet[comoving 5 and 10$h^{-1}$ Mpc box lengths]{may_structure_2021} only reach $z = 3$, after which the de Broglie scale cannot be well resolved. There is however a great deal of progress currently underway in the numerical techniques involved in FDM simulation; recently \cite{chan_cosmological_2025} were able to apply a novel hybrid scheme to successfully perform a zoom-in simulation of a Milky way-sized FDM halo down to $z = 0$. Regardless, FDM simulations are not currently equipped to simulate the large scale structure down to low redshifts.  As a result, it remains unclear to what extent the wave-like features predicted at high redshift would persist into the late-time Universe on the scale of the cosmic web. Nevertheless, the advent of large spectroscopic surveys such as the Sloan Digital Sky Survey \citep[SDSS;][]{york_sloan_2000}, GAMA \citep{driver_gama_2009}, and the Dark Energy Spectroscopic Instrument \citep[DESI;][]{desi_collaboration_desi_2016} has provided statistically rich samples of galaxies in the low-redshift Universe. These data sets have been successfully used to map and characterise the cosmic web, but have not yet been fully exploited in the search for observational signatures of FDM.

Statistical investigations of the large-scale cosmic web involve using galaxy positions as tracers of the underlying matter distribution, using a variety of well-established techniques (e.g. \citealt{libeskind_tracing_2018} and the references within). If the dark matter distribution within filaments exhibits interference fringes, as predicted in FDM models, galaxies may preferentially be found in regions with enhanced dark matter density associated with constructive interference. Whilst galaxies are not expected to trace the dark matter distribution exactly on these scales, owing to baryonic processes and relative motions, even a relatively weak coupling between the galaxy and dark matter distribution could manifest as statistically significant periodic features in the radial distribution of galaxies around filaments. The structure of late-time filaments in an FDM cosmology has not been accurately predicted due to the limitations of existing simulations. Nonetheless, searching for periodicities in the galaxy distribution in filaments offers a promising observational window onto fuzzy dark matter, enabling tests of FDM predictions using large low-redshift galaxy samples and mature cosmic web reconstruction methods.

In this work, we present a novel methodology to investigate whether the distributions of galaxies around cosmic web filaments display statistically significant periodicities. Using the assumption that on filament scales, the galaxy distribution traces the underlying dark matter distribution to some extent, we construct 1-D filament number profiles and measure the distribution of galaxies around filaments. We perform a Fourier analysis of the filament profiles and compute the median power spectra for the filament catalogue of \cite{tempel_detecting_2014} in order to search for characteristic spatial frequencies. To demonstrate the methodology, these measurements are compared to a simple phenomenological model that includes a single periodic component of the form $A\cos(2\pi d/\lambda)$, where $A$ parameterises the amplitude of the periodic term and $\lambda$ its wavelength. By comparing the model and observations using a $\chi^2$ analysis, we identify regions of the parameter space in which the observations are consistent with the presence of such periodic features and those which are disfavoured.

This paper is organised as follows. Section \ref{Sec: Data} describes the filament catalogue and galaxy sample. Section \ref{Sec: Methodology} outlines the methodology, including the calculation of distances to filaments, the form of the model filaments, and the treatment of galaxies in clusters and groups. The measurement of the power spectrum are presented in Section \ref{Sec: Power Spectrum}. The results are presented and discussed in Section \ref{Sec: Results}, and our conclusions and future prospects are given in Section \ref{Sec: Conclusions}.

Throughout this work, we adopt the \textit{WMAP 9} cosmology \citep{hinshaw_nine-year_2012}, with  \textit{H$_0$} $= 69.32 \text{km\,s}^{-1} \text{Mpc}^{-1}$. 

\section{Data} \label{Sec: Data}
In this work, we make use of the filament sample of \cite{tempel_detecting_2014}, which was constructed using the galaxy sample presented in \cite{tempel_groups_2012}. This galaxy sample is largely based on the Sloan Digital Sky Survey DR8 Main Galaxy Sample \citep{york_sloan_2000, aihara_eighth_2011}.

\subsection{Galaxy Sample} \label{Galaxy Sample}

The galaxy sample of \cite{tempel_groups_2012} is based on the SDSS DR8 Main galaxy sample with a small number of modifications. An extinction-corrected apparent magnitude limit of $m_r = 17.77$ is imposed, yielding a uniform sample. Through visual inspection, the authors removed duplicate galaxies and misclassifications. In addition, a lower CMB corrected redshift limit of $z = 0.009$ was applied to exclude the local supercluster. A minor difference exists between the galaxy sample used in \cite{tempel_groups_2012} and that employed by \cite{tempel_detecting_2014} for filament identification: the latter imposes an addition upper CMB corrected distance of 450 $h^{-1}$Mpc, beyond which, the sample becomes too diluted. We also adopt the same limit here, resulting in a final sample of 499,340 galaxies.

\subsection{Filament Sample} \label{Filament Sample}

The filament catalogue of \cite{tempel_detecting_2014} was constructed by applying the Bisous method \citep{stoica_detection_2005, stoica_three-dimensional_2007} to the galaxy sample described above in \autoref{Galaxy Sample}. The Bisous method is a marked-point process that models filaments as a stochastic network of connected cylinders. These cylinders are characterised by parameters such as radius, height and orientation, and collectively trace the filamentary structure of the cosmic web. A detailed description of the Bisous method as well as marked-point processes can be found in \cite{tempel_detecting_2014} and references therein. In total, the filament catalogue contains 15,421 individual filaments.

\subsection{Group and Cluster Catalogue} \label{Group and Cluster Catalogue}

Galaxy clusters and groups can be found inside filaments, these are turbulent, non-linear regimes with galaxy populations that may be influenced by the cluster/group potential, hindering our search for periodicities in filaments.  Considering this, we therefore limit our study to filaments only, where FDM signatures are predicted to be prominent \citep{zimmermann_interference_2024}, excluding areas near galaxy clusters and groups.

To identify galaxies likely to be affected by groups and clusters, we use the galaxy group catalogue of \cite{yang_galaxy_2007}. This catalogue was constructed by applying an iterative halo-based group finder to the New York University Value-Added Catalogue \citep{blanton_new_2005}. A friends-of-friends algorithm first identifies tentative groups and estimates halo properties such as mass, size and velocity dispersion. These estimates then inform updated group memberships, and the procedure iterates until the memberships converge. In this work, we adopt halo masses derived from ranking group luminosities, as provided in the catalogue.

Using these halo properties, we identify galaxies satisfying $|z - z_{h}| < 3z_{\sigma}$, where $z_{\sigma}$ is the redshift interval corresponding to the halo velocity dispersion. For haloes with $10^{13}\text{M}_{\sun} < M_{\text{h}} < 10^{14}\text{M}_{\sun}$, galaxies within a projected distance of $R_{180}$ from the halo centre are removed from the sample\footnote{$R_{180}$ was determined using equation (5) in \cite{yang_galaxy_2007}.}. For more massive haloes ($M_{\text{h}} > 10^{14}\text{M}_{\sun}$), we exclude galaxies within $2.5R_{180}$. This separation corresponds to the commonly adopted distinction between  groups and clusters. This larger exclusion radius for the more massive clusters helps to isolate the filament \citep{rost_span_2021} and is chosen to conservatively remove galaxies potentially influenced by the cluster environment, whilst avoiding excessive dilution of the sample for lower-mass groups (see \citealt{okane_effect_2024}).

We use spectroscopic redshifts queried directly from SDSS rather than the CMB-corrected redshifts provided in the \cite{tempel_groups_2012} catalogue, in order to maintain consistency with the group catalogue. The magnitude of this correction is negligible for our sample (with a maximum difference of 0.00123 in redshift) and does not affect our results.

\begin{figure*}
	\includegraphics[width=2\columnwidth,keepaspectratio]{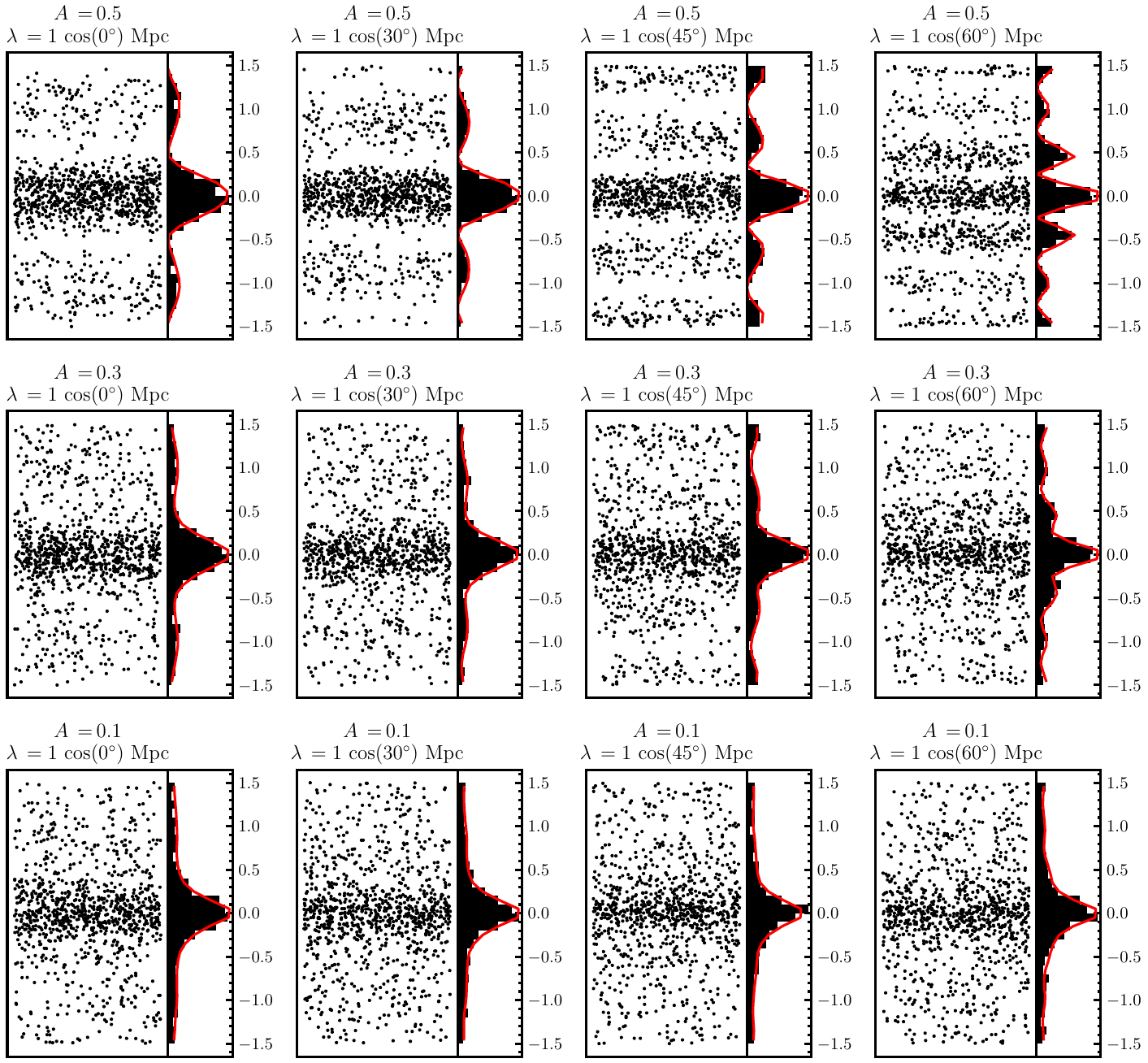}
    \caption{A visualisation of the simple toy model used in this work for various $A$ and $\lambda$, using $\lambda_0 = 1$\,Mpc and varying the inclination angle $\theta$. The 1-D filament profiles are taken and 1,000 distances are drawn from the model (red line atop of the histogram) and translated to positions. It can be seen that decreasing $A$ at a given $\lambda$ reduces the density contrast of the periodicities, whilst increasing the inclination decreases the observed projected wavelength. It is clear from the histograms that variations in $A$ and $\lambda$ produce distinct differences in the filament profiles, especially for larger values of $A$. }
    \label{fig:Visualisation}
\end{figure*}

\section{Methodology} \label{Sec: Methodology}

\subsection{Measuring Possible Periodicity}

In order to determine whether a periodicity exists, we measure galaxy displacements perpendicular to filaments by calculating the 2D projected distance to the nearest point of the nearest filament within a thin redshift slice ($\pm\,0.005$). Reliable distances are crucial for our analysis, and given the substantially larger uncertainty in radial (redshift) positions compared to sky coordinates (right ascension and declination), the 3D distances provided in the filament catalogue are unsuitable for this purpose.

We define galaxy displacements as signed ($+/-$) distances measured perpendicular to the filament spine, assigning positive values to galaxies on one side of the filament in the plane of the sky and negative values to those on the other. This choice is motivated by the fact that the geometric filament spine does not necessarily coincide with a peak or trough of the underlying dark matter density. If a periodic modulation exists with a non-zero phase, the resulting density profile would be asymmetric. Using signed distances preserves the phase information of any periodic signal -- and hence its characteristic wavelength -- even in the presence of such asymmetry. By contrast, using absolute distances would symmetrise the profile and wash out this information.

\begin{figure*}
	\includegraphics[width=2\columnwidth,keepaspectratio]{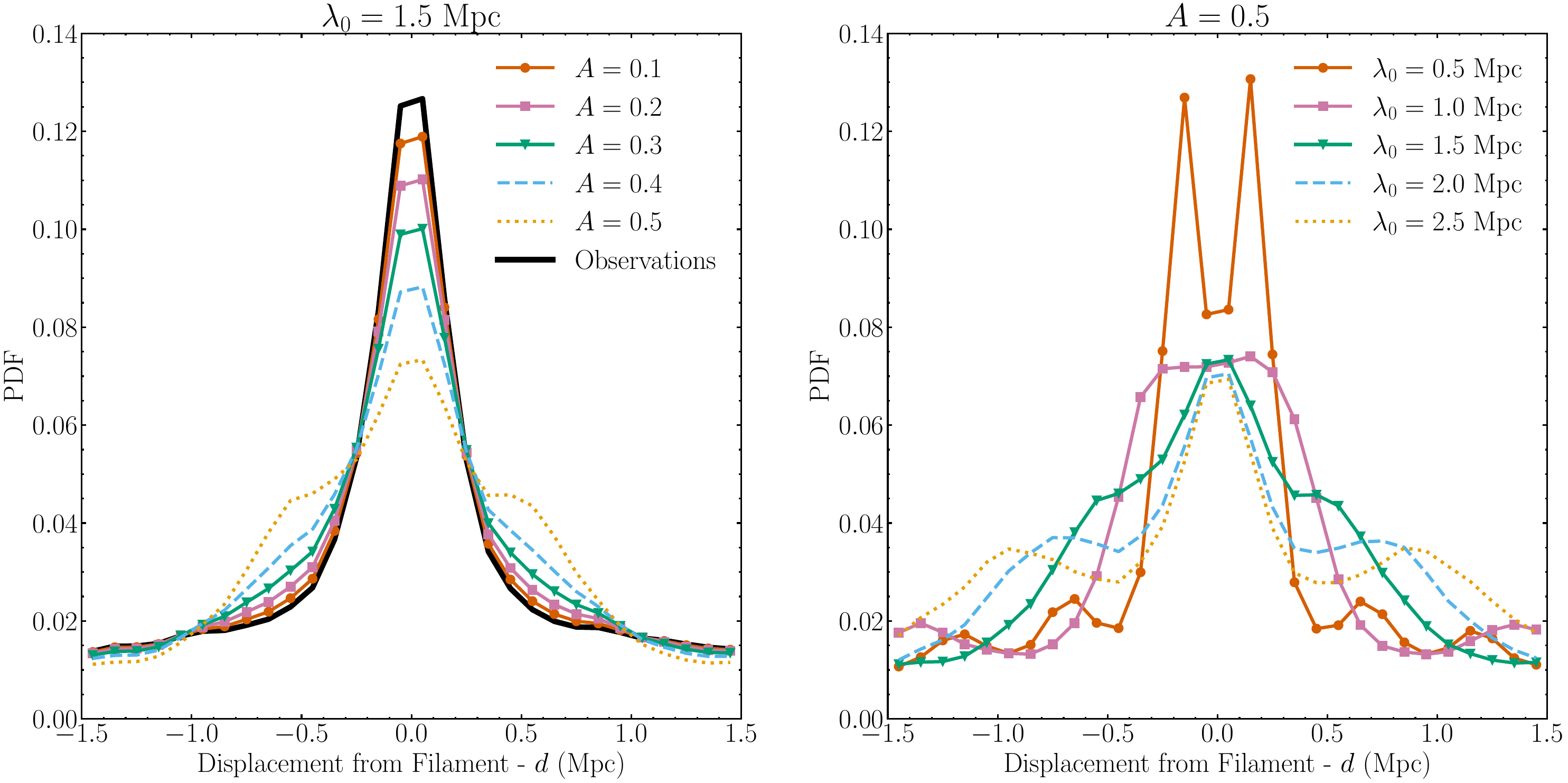}
    \caption{Example of the envelope term $F_{\rm e}$ as function of both $A$ and $\lambda_0$, sampled in increments of 0.1\,Mpc (see \autoref{Sec: Observational Power Spectra}). This function effectively parameterises the shape of the filaments in the absence of periodicity. It can be seen that this shape varies significantly with both $A$ and $\lambda_0$, highlighting the need to properly account for this in the modeling.}
    \label{fig:Waveless}
\end{figure*}

In the catalogue, filaments are represented as networks of connected 3D points, which we treat as continuous structures formed by straight lines connecting these points. We convert the comoving 3D cartesian coordinates ($X$, $Y$, $Z$) provided in the catalogue to sky coordinates ($\alpha$, $\delta$, $z$) and assign each filament a mean redshift computed from its constituent points. For each filament, we select all galaxies with redshfits within $\pm\,0.005$ of the filament redshift. This redshift interval was chosen such that it is large enough to account for redshift space distortions associated with the clusters and groups in the field \citep[Fingers-Of-God effect;][]{jackson_fingers_1972}, whilst also ensuring that even the largest of filaments can be treated appropriately regardless of their orientation; we want to include all filaments including those not perpendicular to the plane of the sky. Displacements are then calculated using the angular diameter distance corresponding to the mean redshift of the filament. We include only galaxies located within the filament cylinder; galaxies just outside the cylinder -- such as those near the filament start or ends -- are excluded.

To avoid double-counting, each galaxy is assigned only to the closest filament. Excluding galaxies assigned to groups and clusters (cf. \autoref{Group and Cluster Catalogue}), our final sample comprises of 214,173 unique galaxies within 10\,Mpc of a filament, of which 122,288 lie within 1.5\,Mpc of a filament. The number of filaments as a function of galaxy members (cf. \autoref{Sec: Observational Power Spectra}) is shown in \autoref{fig:FilamentMembers}, we identify 4,394 filaments containing at least 10 member galaxies within 1.5\,Mpc.

\subsection{Model and Assumptions}

Our most significant assumption moving forward is that, on filament scales, galaxies trace the underlying dark matter distribution to some extent. In other words, if there are periodicities in the dark matter, corresponding periodicities may also appear in the galaxy distribution around filaments. Over large scales, it is well established that the galaxy distribution, whilst biased, does trace the underlying dark matter, and this fact underpins observational modeling of the cosmic web \citep{kaiser_spatial_1984,tegmark_time_1998,libeskind_tracing_2018}. However, on smaller, intra-filamentary scales,  the degree of this coupling is uncertain and may break down. Even in a Lambda CDM framework, there is still uncertainty, for example, \cite{watson_extreme_2012} investigate the radial distribution of galaxies in their host halos using SDSS data and find that the distribution of faint galaxies is consistent with a Navarro--Frenk--White profile, but the more luminous galaxies are not, implying that more luminous galaxies are poorer tracers of the underlying dark matter on small scales. Whereas \cite{shin_spatial_2022} compare the spatial distribution of the dark matter and baryonic components in the IllustrisTNG cosmological hydrodynamical simulation and conclude that in and around galaxy clusters, galaxies are still a reliable tracer of the dark matter distribution. It is clear that the jury is still out here, especially for FDM where the lack of adequate simulations means it remains unclear whether the dark matter distribution is well represented in the galaxy distribution. Under the assumption that the galaxy distribution does trace the dark matter to some extent, even a weak correlation could be measurable using a statistically rich galaxy sample and robust methodology.

There is no clear expectation for the shape of these periodicities in the galaxy distribution. Simulations are generally restricted to early cosmic times, and it is not obvious how such interference features would evolve over time. To allow us to compare our observations, we model filaments using a simple model, examples of which are shown in \autoref{fig:Visualisation}. For simplicity, we adopt a model consisting of a envelope term, $F_{\rm e}$, describing the overall shape of the filament, and a single periodic term. Two parameters are left free: $A$, the maximum density contrast(amplitude) of the periodic term, which characterises the strength of the periodicities, and $\lambda$, the wavelength of the periodicities. As a function of displacement to filaments $d$ ($+/-$ distances), the model is written as
\[F(d) = F_{\rm e}(d)\left[1-A + A \cos \left( \frac{2\pi d}{\lambda}\right)\right].  \tag{1} \label{FilamentProfile}\]
The $1-A$ term is such that the model is normalised to a maximum value of 1. We model these interference patterns as 2D planes embedded within the larger filament structure, inclined at an angle $\theta$ to the line of sight. Consequently, the observed periodicity depends on the inclination:
\[\lambda = \lambda_0 \cos{\theta},  \tag{2} \]
where $\lambda_0$ represents the maximal (face-on) wavelength.
In \autoref{fig:Visualisation}, we visualise the spatial distribution of galaxies tracers for different model parameters. It is clear that as $A$ decreases, the contrast between the troughs of destructive interference and the peaks of constructive interference decreases. It is further clear how an increase in the inclination, which is equivalent to a decrease in $\lambda$, reduces the wavelength of the observed projected periodicities.

Whether this model accurately describes FDM is not known. However, an assumption about the profile of filaments is necessary to compare to the observations and our assumption is strengthened by the theoretical work of  \citet[see, in particular, their figure 1]{zimmermann_interference_2024}, which employs an analytical treatment of an FDM filament, predicting a degree of cylindrical asymmetry consistent with this approach. A more detailed or sophisticated model would require large scale simulations at low redshift, which is not currently possible.

We further assume that all filaments can be described by our model with a single value of $A$ and $\lambda_0$. There is extensive evidence that filament structural parameters depend on various factors, such as environment and proximity to clusters (e.g. \citealt{pimbblet_intercluster_2004,gonzalez_automated_2010,cautun_evolution_2014, kim_large-scale_2016, kuchner_mapping_2020,rost_span_2021, wang_boundary_2024,zhang_statistical_2024}). Given the uncertainties involved, attempting to account for these dependencies is not practical.

Our primary goal is to search for periodicities in the filament distribution, which is independent of the model assumptions; the model only informs the interpretations and demonstrates the method. This methodology could also be applied to alternative models of filament structure where a degree of periodicity is potentially present.

We stress that, although $\lambda_0$ does correspond to the wavelength of the periodicity, it should not be directly interpreted as the de Broglie wavelength of the FDM particle. We refrain from claiming that this periodicity corresponds to the de Broglie wavelength, given the substantial uncertainties associated with the structure of the interference fringes. For example, filaments are generally expected to evolve over time owing to the expansion of the Universe and other physical processes; it is therefore unclear whether such interference patterns would evolve commensurately or instead remain fixed in scale. These uncertainties, together with the assumptions inherent in our analysis, preclude a definitive association. This is not to suggest that the observed periodicity does not have its origin in the de Broglie wavelength, but rather that the information available at the later times probed by the observations is insufficient to establish a direct connection. Consequently, we restrict our analysis to identifying periodicities, rather than attempting to constrain the de Broglie wavelength (and hence the particle mass).

\subsubsection{Modelling the Underlying Filament Envelope} \label{Sec: Envelope}

The envelope term $F_{\rm e}(d)$ parameterises the underlying shape of the filaments in the absence of periodicity. Choosing an appropriate form for $F_{\rm e}(d)$ is crucial, as it directly influences the resulting power spectrum. The existing literature offers little guidance here, as there is no clear consensus on the expected filament shape, and any choice introduces a degree of model dependence. Therefore, we adopt a model-agnostic, purely empirically form for $F_{\rm e}(d)$. Examples of $F_{\rm e}(d)$ for varying $A$ and $\lambda_0$ are shown in \autoref{fig:Waveless}.

We impose the requirement that, irrespective of the filament form, the stacked profile of all the filaments must match observations. Within our model, this boundary condition can be expressed as
\[ \sum[F(d)] = \int_0^{\frac{\pi}{2}} F_{\rm e}(d)\left[1-A + A \cos \left(\frac{2\pi}{\lambda_0\cos(\theta)}\right)\right]f(\theta) d\theta ,\tag{3}\]
where $f(\theta)$ is the probability distribution function of the filament inclination, which is uniform with $f(\theta) = 2/\pi$. Since the envelope term represents the filaments intrinsic shape, we assume cylindrical symmetry, implying that $F_{\rm e}(d)$ is independent of $\theta$. This allows us to express the envelope function as
\[F_{\rm e}(d)= \frac{\sum[F(d)]}  {\int_0^{\frac{\pi}{2}} \frac{2}{\pi} \left[1-A + A \cos \left(\frac{2\pi}{\lambda_0\cos(\theta)}\right)\right] d\theta}. \tag{4} \label{EnvelopeEq}\]
This formulation ensures that, for any choice of $A$ and $\lambda_0$, the envelope term multiplied by the periodic term, averaged over random inclinations, reproduces the observed stacked filament profile. The numerator in \autoref{EnvelopeEq} corresponds to the observed stacked profile, whilst the denominator can be evaluated numerically. In the left plot of \autoref{fig:Waveless}, $A$ varies (for a fixed $\lambda_0$), in the right plot, $\lambda_0$ varies (for a fixed $A$). It is evident that the shape of the model filaments varies significantly with these parameters. These differences will lead to differences in the power spectra and must be taken into account when constructing a model to properly compare to observations.

\section{The Power spectra} \label{Sec: Power Spectrum}

In this section, we describe our methodology for obtaining the power spectra from observations, and our simulated model power spectra.

\begin{figure*}
	\includegraphics[width=2\columnwidth,keepaspectratio]{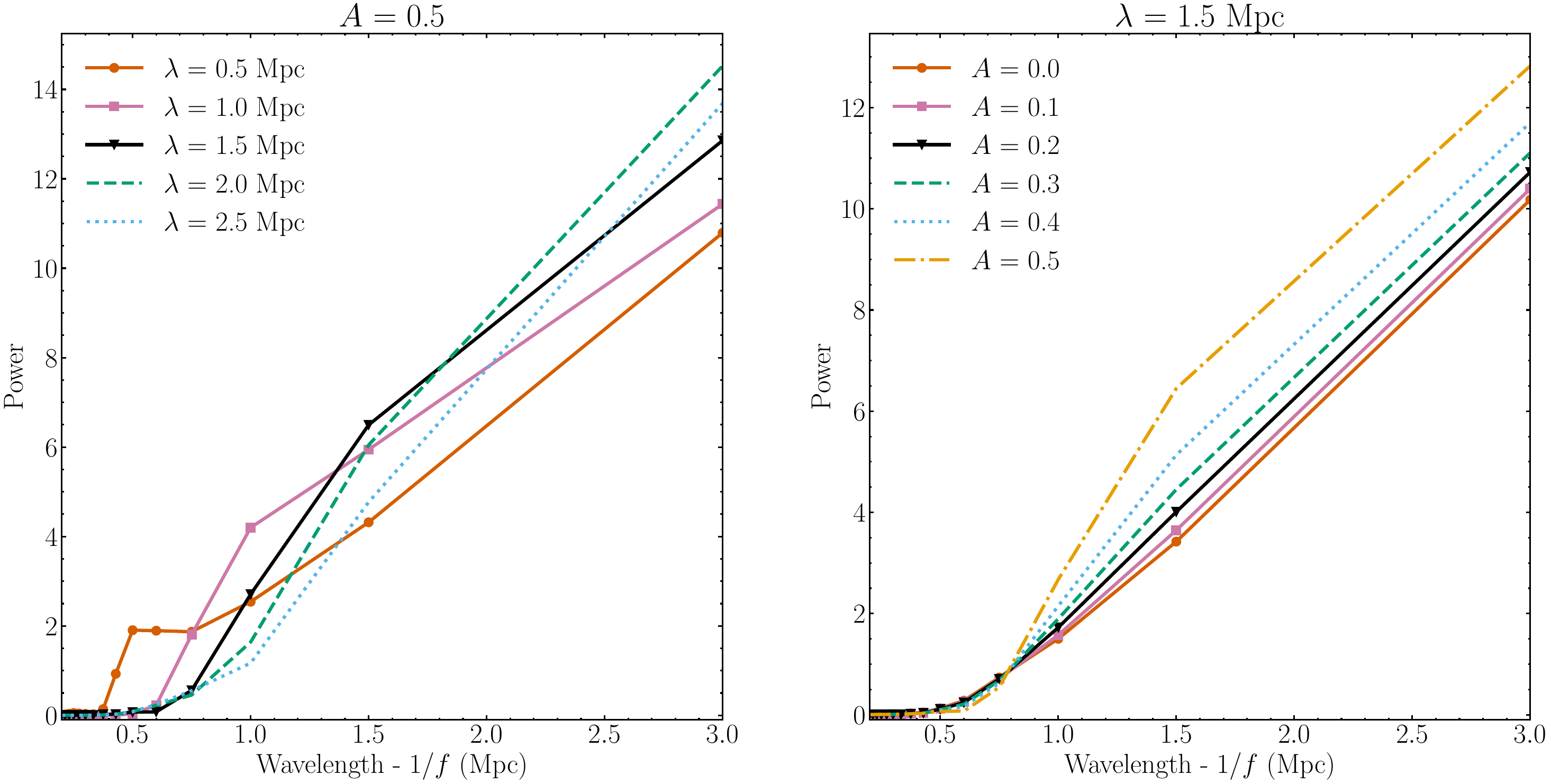}
    \caption{Examples of the power spectra for idealised simple model versions of filaments for different values of $A$ and $\lambda_0$ for constant $\lambda$ (Left) and constant $A$ (Right). The filaments are constructed from 1,000 mock galaxies with 250 noise estimates, repeated a total of 250 times. It is clear from these plots that both differences in $A$ and $\lambda$ produce clear and measurable differences in the power spectra.}
    \label{fig:ExamplePS}
\end{figure*}

\subsection{The observed power spectra} \label{Sec: Observational Power Spectra}

For each filament, we construct filament profiles by taking a histogram of galaxy displacement from the filament axis. In \autoref{fig:FilamentMembers}, we present the number of filaments as a function of filament members within $1.5\,$Mpc. We find that imposing a minimum member limit of 10 galaxies leaves us with a sample of 4 394 filaments, providing a statistically robust dataset whilst excluding filaments with too few members to produce reliable power spectra.

The choice of both the profile length (radial extent) and histogram bin size directly affects the measurable frequency range of our power spectra. In particular, periodic components with frequency larger than half the sample frequency (the Nyquist Frequency; \citealt{nyquist_certain_1928,shannon_communication_1949}) cannot be accurately measured. Frequencies above this limit can lead to aliasing, in which power from high frequency components is incorrectly measured at lower frequencies. To mitigate this, we must balance adequate sampling with avoiding oversampling; too small a bin size may fail to capture the filament structure accurately. We adopt a bin size of 0.1\,Mpc as a reasonable compromise.

Similarly, we cannot measure frequencies smaller than the Rayleigh frequency/criterion, representing the minimum frequency measurable in Fourier analysis for a signal of finite length, defined as the reciprocal of the profile length. Therefore, the profile length should be maximised. Previous studies indicate that filaments tend to have a characteristic width on the order of $\sim 1\,$Mpc \citep{colberg_inter-cluster_2005,aragon-calvo_multiscale_2010,bond_crawling_2010,gonzalez_automated_2010,cautun_evolution_2014,kuchner_mapping_2020,Castignani_Virgo_1,wang_boundary_2024,aguerri_galaxy_2026}, and extending the profile much beyond this risks including galaxies not associated with the filament. Based on this, we limit our filament profiles to displacements within $\pm\,1.5\,$Mpc, yielding a measurable wavelength range of 0.2--3.0\,Mpc.

We determine the power spectra of filament profiles via the Fourier transform of the autocorrelation function. By the Wiener--Khinchin theorem \citep{wiener_generalized_1930}, this is equivalent to the square of the Fourier transform and is generally preferred with noisy data.

To suppress edge effects, we apply a Hanning window \citep{blackman_measurement_1958} to the filament profiles before computing the autocorrelation function and Fourier transform. The Hanning window is a tapering function of the form $\propto \cos^2(x)$, bringing the profile to 0 at the start and end to reduce discontinuities. The resulting power spectra are normalised by $(N^2 \delta d)$ to remove dependence on the profile sampling $\delta d$ and the number of galaxies in each filament $N$.

Filament profiles in this work typically contain $\sim\,10$ galaxies (\autoref{fig:FilamentMembers}). Consequently, the measured power spectra are dominated by shot noise, which we expect to be Poisson in nature and approximately white, with a constant contribution across all frequencies. To correct for this, we explicitly model the noise rather than leaving it as a free parameter. For each filament profile, we generate 250 mock noise profiles by drawing random values from a Poisson distribution scaled by $N_{\text{bin}}$ and compute the median power spectra of these noise realisations. This median is then subtracted for the measured filament power spectrum to produce noise-corrected results.

We then compute the median power spectra across the filament sample. We use the median rather than the mean because the power spectrum is a non-linear statistic, and outliers -- such as noise erroneously identified as filaments -- could disproportionately affect the mean.

To estimate uncertainties, we preform bootstrapping with replacement. From the 4 394 filaments, we draw resampled filaments sets and repeat the median power spectrum calculation a total of 250 times. The final median power spectrum is taken as the $50^{\text{th}}$ percentile of the resulting distribution, and the $1\sigma$ uncertainties are defined as half the difference between the $84.1$ and $15.9$ percentiles.

\subsection{The simulated power spectra}

To compare with the median power spectra from observations, we generate simulated filaments. Here, a simulated filament refers to a set of random galaxy displacements drawn from the model described in \autoref{FilamentProfile}, treated as a probability distribution function. For each of the 4,394 filaments in our sample, we generate a simulated filament with the same number of galaxy members as the observed filament. 

We determine the median power spectra of the simulated filaments using the same methodology described in \autoref{Sec: Observational Power Spectra}. One distinction concerns the estimation of uncertainties. Unlike the observational data, where errors are obtained via bootstrapping with replacement, here we repeat the full process, including the random drawing of displacements to filaments, a total of 250 times. The final simulated median power spectrum is taken as the $50^{\text{th}}$ percentile of this distribution and the $1\sigma$ uncertainty as half the difference between the 84.1 and 15.9 percentiles.

In \autoref{fig:ExamplePS}, we present examples of the power spectra for a single simulated filament containing 1,000 galaxies. This is substantially larger than any individual filament in our sample but illustrates that variations in $A$ and $\lambda$ produce significant, measurable differences in the power spectra. 

To interpret the shape of the power spectra seen here, we must first understand what contributes to it. There are two factors in play, the first is the periodic cosine component in \autoref{FilamentProfile}, and the second is the overall shape of the filament profile (envelope) in the absence of a periodic component, described by the envelope term $F_{\rm e}$. The general trend observed in \autoref{fig:ExamplePS} is that at a given $\lambda$, the power increases with wavelength (right plot); this characteristic shape is due to the overall shape of the filaments. In contrast, when the wavelength $\lambda$ is varied at a fixed $A$ (left plot), distinct features emerge in the power spectra. For example, at $\lambda = 0.5$, a peak appears at 0.5 Mpc, while the overall trend of increasing power with wavelength is preserved. In other words, the power spectra for a given model filament can loosely be thought of as the power spectra of the envelope term, with an additional peak in the spectra at wavelengths corresponding to $\lambda$. This simplified description can also be derived analytically, the results of which are presented in appendix \ref{Anayltical Treatment}.

\begin{figure*}
	\includegraphics[width=2\columnwidth,keepaspectratio]{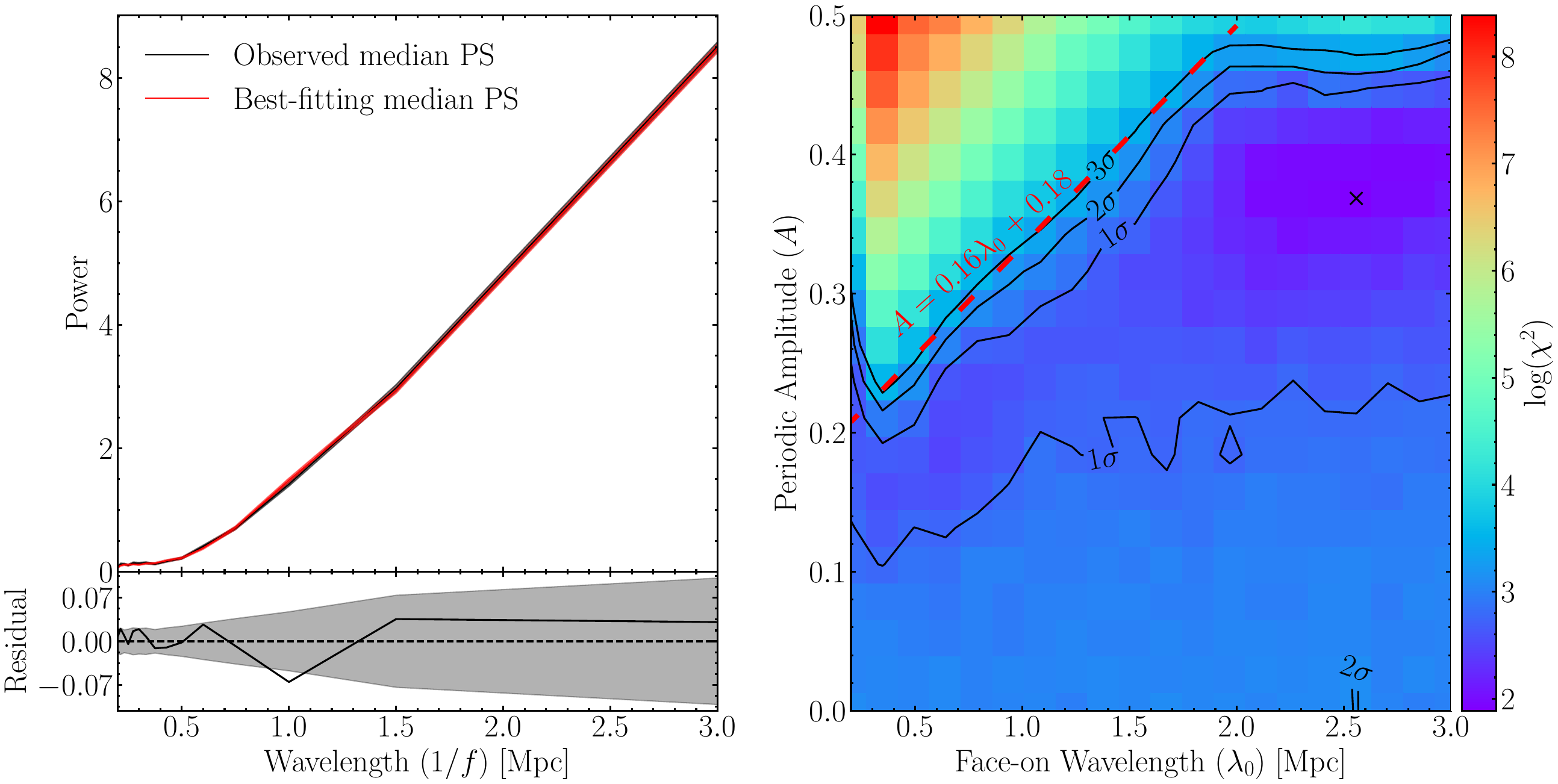}
    \caption{\textbf{(Left)} The observed median power spectra (black) compared to the best fitting simulated median power spectra (red) along with their respective $1\,\sigma$ error of the difference. The residual plot is shown below in the black solid line, with the shaded grey region representing the $1\sigma$ confidence interval in which they agree. \textbf{(Right)} The parameter space of $A$ and $\lambda_0$ coloured by the $\chi^2$ of the corresponding model. The black lines denote the 1, 2 and 3\,$\sigma$ contours and the black cross represents the best fitting $A$ and $\lambda_0$ for which the $\chi^2$ is minimal. The red dashed line traces the approximately linear regime of the $3\,\sigma$ countour, above which we exclude values at the $3\,\sigma$ level. It is clear that the vast majority of the parameter space is consistent with the best fitting model (within $2\,\sigma$), including all models with $A = 0$. Suggesting that the distribution of galaxies around filaments is consistent with no periodicities.}
    \label{fig:SimToData}
\end{figure*}

\section{Results and Discussion} \label{Sec: Results}

We compare the median observed power spectra to the median simulated power spectra for values of $A$ and $\lambda_0$. The parameter space is mapped using the goodness-of-fit parameter $\chi^2$ (\autoref{fig:SimToData}, right side), and $\chi^2$ values are converted to 1, 2 and 3\,$\sigma$ contours using the $\chi^2$ distribution with $15/1.5 - 2 = 8$ degrees-of-freedom. This corresponds to 15 Fourier freuency measurements, corrected for the equivalent noise bandwidth of the Hanning window \citep[factor of 1.5;][]{harris_use_1978}, and 2 fitted parameters. It is evident from \autoref{fig:SimToData} that the majority of the parameter space is consistent with the observations (within $2\sigma$). Crucially, we find that all the models with $A = 0$ are within the $2\sigma$ contours and are statistically consistent with the observations at the $2\sigma$ level. In other words, our results suggest that the observations are consistent with a simple model of this form with no periodicity. It is clear however that given the size of the $2\sigma$ contours we can make no strong claim for which values of $A$ and $\lambda_0$ best match observations. However, we find that there is a region of the parameter space which is heavily disfavoured and can be used to exclude specific values of $A$ and $\lambda_0$. To quantify this, we fit a straight line to the $3\,\sigma$ contour in the range [0.35, 1.80] Mpc, where the contour is approximately linear (red dashed line). Based on this linear fit we can exclude values of $A$ for which $A > 0.16\lambda_0 + 0.18$, at the 3\,$\sigma$ level for $0.2\,\text{Mpc}\,\lesssim \lambda_0 \lesssim 2\,\text{Mpc}$. We note that even with a simple model such as that adopted in this work, searching for periodicities in the galaxy distribution around filaments can produce constraints on model parameters.

Even though we can rule out some values of $A$ with our results, it is clear there is a broad range of $A$ and $\lambda_0$ values consistent with observations. In particular, the conclusion that the observations are consistent with the $A = 0$ models is interesting but not straightforward to interpret. This is due to the uncertainties in the underlying assumptions. As a reminder, a key assumption we made was that, on the scales probed here, galaxies trace the dark matter distribution. The validity of this assumption, especially in a FDM universe is not well established due to the lack of sufficient FDM simulations for the present-day Universe. Consequently, periodicities may exist in the dark matter profiles of filaments but may not be reflected in the galaxy distribution to a measureabale extent. Without the necessary FDM simulations however, inferences are therefore limited. A recent relevant study is that of \cite{kousha_nearest_2023}, which compares FDM and CDM simulations. Through a nearest neighbour analysis, they find significant differences in the spatial distributions of dark matter haloes between FDM and CDM, with FDM haloes showing reduced clustering. The authors suggests that such analyses of galaxy positions can offer valuable insight into the effects of FDM, further reinforcing the value of studies such as this.

Another assumption made in this work is that all filaments can be represented by a single value of $A$ and $\lambda_0$, neglecting any dependence on filament properties such as length, mass, curvature etc. Whilst we acknowledge that this dependence could lead to a departure from our model, attempting to account for such dependences is far beyond the scope of possibilities without sufficient simulations.

An implicit assumption made in this work is that the filaments extracted using the galaxy distribution well represent the dark matter distribution of filaments. \cite{zakharova_filament_2023} investigated how different tracers produce different filament networks by applying the DisPerSE filament finder \citep{sousbie_persistent_2011} to the results of the GAlaxy Evolution and Assembly (GAEA) semi-analytic model coupled to the Millennium simulation, finding that although filaments extracted with galaxies and those with dark matter tend to agree, they do not completely. This suggests that filaments extracted using galaxies may not be entirely representative of the underlying dark matter distribution. Whether this is also true in a FDM cosmology is something that would also require large scale simulations to test.

We further highlight the removal of cluster/group galaxies from our galaxy sample. Clusters and groups are not randomly distributed in filaments and tend to be found near filament spines \citep{tempel_detecting_2014}. Removing the galaxies in groups and clusters would systematically remove galaxies closer to the filament spine and could introduce dips in the profiles and smear out any periodicities present. Again, we stress that the model employed here is a simplistic, idealistic model due to the vast uncertainties from the lack of sufficient FDM simulations.

Whilst our study employs a large sample of filaments, our analysis may be limited by the number of filament members. Although the number of filaments is relatively large (4,394), most filaments contain $\sim\,10$ galaxies (\autoref{fig:FilamentMembers}), potentially insufficient to detect periodicities. This likely contributes to the vast portion of the parameter space contained within the $2\sigma$ contour. This limitation arises from the wide but relatively shallow SDSS Main galaxy sample ($r \leq 17.77$). Future studies, such as the 4MOST Hemisphere Survey \citep{taylor_4most_2023} and the DESI Bright Galaxy Survey \citep{hahn_desi_2023}, will provide deeper, larger samples ($r \leq 20$), vastly increasing the number of galaxies per filament which could improve the constraints in the parameter space.

In this work, we utilise galaxy distributions around filaments. A promising alternative, as highlighted by \cite{zimmermann_interference_2024}, could be to search for FDM signatures via filament surface density maps obtained through gravitational lensing. For example, \cite{hou_probing_2025} use gravitational lensing observables to probe dark matter substructure in mock catalogues, excluding FDM particle masses below $\sim10^{-22}$ eV. There is potential in using lensing to probe dark matter substructure, including filamentary structure (e.g. \citealt{hyeonghan_weak-lensing_2024,scognamiglio_ultra-high-resolution_2026}) and circumvent the need for galaxy tracers entirely. Nevertheless, using galaxy positions remains valuable, providing a relatively unexplored observational dataset, over a large volume, complementary to existing studies.

In the broader context of FDM, simulations generally predict structural periodicities on scales comparable to the de Broglie wavelength \citep{schive_cosmic_2014,mocz_first_2019,li_numerical_2019,mocz_galaxy_2020,may_structure_2021, kulkarni_if_2022}. Although this behaviour is observed at the high redshifts probed by current simulations on cosmological scales, it remains unclear whether it persists at the lower redshifts accessible to observations of large-scale structure. This uncertainty arises primarily from the lack of information available at late times and from unresolved questions regarding the evolution of filamentary interference patterns. Given these uncertainties, we focus on developing robust tools and methodologies for detecting FDM-induced periodicities. This work therefore complements existing searches for observational signatures of FDM, while deliberately avoiding explicit constraints on the FDM particle mass. 

Nevertheless, it is informative to consider the implications of interpreting the measured periodicities as corresponding directly to the de Broglie wavelength. Under this assumption, we can comment on the range of FDM particle masses probed by this work. To avoid distracting from the primary aims of the paper, we defer this discussion to Appendix \ref{FDM Masses}.

Within this broader context, it is important to note that our analysis is restricted to the canonical single-component FDM model originally proposed by \cite{hu_cold_2000}, which is represented by a single FDM species. More complex realisations of FDM, such as multi-component models (e.g. \citealt{medvedev_cosmological_2014,todoroki_dark_2022}) or models with self-interactions (e.g. \citealt{capanelli_cold_2025, lopez-sanchez_core-halo_2025}), may require fundamentally different assumptions regarding the distribution of characteristic length-scales. Exploring such models lies beyond the scope of the present work and would likely require substantially more sophisticated FDM simulations than are currently available.

Despite these limitations, our analysis demonstrates that a large statistical sample of low-redshift galaxies, combined with well-developed cosmic web extraction techniques, can provide a novel observational probe of fuzzy dark matter. This approach can be further refined and built upon as simulations and theoretical understanding of FDM develop.

\section{Conclusions and future work} \label{Sec: Conclusions}

In this work, we present a novel method to search for signatures of Fuzzy dark matter using the rich dataset of galaxy positions in the nearby universe ($z < 0.2$). We search for periodicities in the distribution of galaxies around cosmic web filaments, motivated by the possibility that the dark matter distribution within filaments may exhibit interference-driven structure in an FDM cosmology, which may manifest in the galaxy distribution. Using a Fourier-based analysis, we compare the observed galaxy distribution around filaments to a simple model of periodic filamentary structure to constrain the model parameters. We find that a majority of the explored parameter space is consistent with observations (within $2\sigma$), including all models with $A = 0$ (no periodicities). We find that values of $A > 0.16$$\lambda_0 + 0.18$ for $0.2\,\text{Mpc}\,\lesssim \lambda_0 \lesssim 2\,\text{Mpc}$ are heavily disfavoured by observations and are excluded at the $3\sigma$ level, demonstrating the capability of this methodology to test and constrain different models.

Given the uncertainties surrounding the structure of FDM filaments at late times, as well as the assumptions inherent in this analysis, the absence of a measurable signal cannot rule out an FDM cosmology. Rather, we conclude that if the universe is well described by a simple FDM model like the one used in this work, any periodicities present in the dark matter distribution of filaments are not yet detectable at a measurable level using galaxy positions alone. 

This work demonstrates that, given a model for FDM filament structure, it is possible to search for FDM signatures within cosmic filaments using galaxy data. This represents a novel application of a relatively unexplored observational dataset that contains a wealth of statistical information. While we make no inference regarding the mass of the FDM particle, this approach provides a complementary probe to existing tests of the FDM paradigm and can be readily adapted as theoretical models and numerical simulations improve.

The primary aim of this work is the identification of FDM-induced periodicities rather than the direct constraint of the FDM particle mass. If the measured periodicities are interpreted as manifestations of the de Broglie wavelength, they imply sensitivity to a particular range of FDM particle masses. We discuss this interpretation in Appendix \ref{FDM Masses}.   

Looking ahead, future spectroscopic surveys such as the DESI Bright Galaxy Survey and the 4MOST Hemisphere Survey will provide substantially larger and deeper galaxy samples, yielding filaments with significantly higher galaxy membership. Such datasets will allow this methodology to move beyond the noise-dominated regime encountered for individual filaments in the present work. An improved statistical sample will be crucial in determining whether periodic signatures in the galaxy distribution around filaments can be detected observationally, and will enable tighter constraints on the relevant parameter space, including sensitivity to smaller wavelengths and amplitudes.

\section*{Acknowledgements}

We thank the anonymous referee for their helpful comments which greatly improved the quality of this work. The authors thank Prof. Jamie Bolton whose comments helped to improve the quality of the paper. 

AAS thanks Tom Broadhurst for many stimulating discussions on Fuzzy Dark Matter.

This work was supported by the Science and Technology Facilities Research Council (grant number ST/X508639/1). AAS, MEG, and UK acknowledge financial support from the UK Science and Technology Facilities Council (STFC; grant ref ST/T000171/1).

Funding for SDSS-III has been provided by the Alfred P. Sloan Foundation, the Participating Institutions, the National Science Foundation, and the U.S. Department of Energy Office of Science. The SDSS-III web site is http://www.sdss3.org/.

SDSS-III is managed by the Astrophysical Research Consortium for the Participating Institutions of the SDSS-III Collaboration including the University of Arizona, the Brazilian Participation Group, Brookhaven National Laboratory, Carnegie Mellon University, University of Florida, the French Participation Group, the German Participation Group, Harvard University, the Instituto de Astrofisica de Canarias, the Michigan State/Notre Dame/JINA Participation Group, Johns Hopkins University, Lawrence Berkeley National Laboratory, Max Planck Institute for Astrophysics, Max Planck Institute for Extraterrestrial Physics, New Mexico State University, New York University, Ohio State University, Pennsylvania State University, University of Portsmouth, Princeton University, the Spanish Participation Group, University of Tokyo, University of Utah, Vanderbilt University, University of Virginia, University of Washington, and Yale University. For the purpose of open access, the authors have applied a Creative Commons attribution (CC BY) licence to any Author Accepted Manuscript version arising. The authors contributed to this paper in the following ways: CJO, UK, AAS, and MEG formed the core team. CJO analysed the data, produced the plots, and wrote the paper along with UK, AAS, and MEG

\section*{Data Availability}

The data underlying this article were accessed from SDSS DR8 http://www.sdss3.org/dr8. The derived data generated in this research will be shared on request to the corresponding author.



\bibliographystyle{mnras}
\bibliography{references} 




\appendix

\section{The analytical power spectrum} \label{Anayltical Treatment}

In this section, we consider the power spectrum for a single model filament described by \autoref{FilamentProfile}.

By the convolution theorem, the Fourier transform of \autoref{FilamentProfile} is equivalent to the convolution ($\ast$) of the Fourier transforms of the envelope and periodic terms.
\[\mathcal{F}[F(d,\theta)] =  \mathcal{F}[F_{\rm e}(d)] \ast \mathcal{F}\left[1-A + A\cos\left(\frac{2\pi d}{\lambda_0 \cos\theta}\right)\right]. \tag{5}\]
The Fourier transform of the envelope term, $\mathcal{F}[F_{\rm e}(d)]$, can only be solved numerically. The Fourier transform of the periodic term, however, can be expressed analytically as 
\begin{align*}
    \mathcal{F}\left[1-A + A\cos\left(\frac{2\pi d}{\lambda \cos\theta}\right)\right] 
    &= (1-A)\delta(\omega) + A\pi\Bigl(\delta\left(\omega - \frac{2\pi}{\lambda_0\cos\theta}\right)\\ &\quad + \delta\left(\omega + \frac{2\pi}{\lambda_0\cos\theta}\right) \Bigr), \tag{6}
\end{align*}
where we employ the sifting property of the Dirac delta function $\delta(x)$\footnote{$\int_{-\infty}^\infty\phi( y)\delta(x - y) dy = \phi(x)$}. Consequently, the complete Fourier transform can be expressed as a linear sum of frequency-shifted copies of $\mathcal{F}[F_{\rm e}(d)]$. Defining the power spectrum as $P(\omega) = \vert\mathcal{F}(\omega)\vert^2$, the power spectrum of \autoref{FilamentProfile} can be written as 

\begin{align*}
    P(\omega) &= (1-A)^2\vert{Q_0}\vert^2 + A^2\pi^2(\vert{Q_-}\vert^2 + \vert{Q_+}\vert^2 + \overline{Q_-}Q_+ + \overline{Q_+}Q_-)\\ 
    &\quad+ A\pi(1-A)(\overline{Q_0}Q_- + \overline{Q_0}Q_+ + Q_0\overline{Q_-} + Q_0\overline{Q_+}),\tag{7} \label{Analytical PS}
\end{align*}
where $Q_0 = \mathcal{F}[F_{\rm e}](\omega)$, and $Q_\pm = \mathcal{F}[F_{\rm e}](\omega \pm2\pi/\lambda_0\cos\theta)$.
Barred quantities denote complex conjugates. 

It is clear that the resulting power spectrum is a combination of the Fourier transforms of the envelope function shifted in frequency.

Moving any further is not entirely straightforward, $\mathcal{F}[F_{\rm e}(d)]$ can only be determined numerically, as $F_{\rm e}(d)$ is empirically determined and lacks a function form. Numerical Foruier transforms are analysed using discrete, finite-range Fourier transforms, as such, the resulting numerical result is valid over only a finite domain [$f_{\rm Rayleigh},f_{\rm Nyquist}$]. Consequently, frequency-shifting (i.e. $\mathcal{F}[F_{\rm e}](\omega \pm2\pi/\lambda_0\cos\theta)$) is not possible without making assumptions about its form outside this domain. 

These issues are of course avoided by choosing a form for the shape of the filaments, however, given the possibility of introducing model dependent scalings, we refrain from doing so in this work.

We also highlight that comparisons to observations would further be complicated by effects such as aliasing and frequency-folding, effects of which are not included in the analytical description described here. 

\section{The measured periodicity and the FDM particle mass} \label{FDM Masses}

We begin by reiterating that the relationship between the characteristic periodicity $\lambda_0$ measured in this work and the de Broglie wavelength of FDM at low redshift remains uncertain. This uncertainty arises both from the lack of appropriate simulations at late times and from unresolved questions regarding how the measured periodicity related to the underlying dark matter.

If one assumes that the measured periodicity $\lambda_0$ is directly equivalent to the FDM de Broglie wavelength, then the wavelength range investigated in this work can be translated into a corresponding range of particle masses. Using equation~1 of \cite{hui_wave_2021}, which expresses the de Broglie wavelength as a function of particle mass $m$ and velocity $v$,

\[\lambda_{\rm dB} = 0.48\,\text{kpc}\,\left(\frac{10^{-22}\,\text{eV}}{m}\right)\left(\frac{250\,\text{km\,s}^{-1}}{v}\right), \tag{8} \label{Eq. Hui 1}\]
and adopting a characteristic velocity of $v = 250$km\,s$^{-1}$, comparable to the infall velocity of dark matter onto filaments \citep{rost_span_2021}, the wavelength range $0.2$\,Mpc -- $3$\,Mpc corresponds to an FDM mass range of $2.4\times10^{-25}$eV -- $1.6\times10^{-26}$eV.

This mass range lies substantially below the canonical FDM mass scale of $\sim10^{-22}$\,eV and overlaps the parameter space that is already strongly constrained by other observational probes (e.g. \citealt{gonzalez-morales_unbiased_2017,irsic_first_2017,rogers_strong_2021,liu_constraining_2025,liu_joint_2025, nadler_warm_2026}). Such constraints are inherently model-dependent and are subject to their own systematic uncertainties. Independent observational tests therefore remain valuable, both as consistency checks and as probes of complementary aspects of FDM phenomenology.

In \autoref{Sec: Results}, we report $3\sigma$ bounds on $A$ as a function of $\lambda_0$. By transforming $\lambda_0$ into an FDM particle mass using \autoref{Eq. Hui 1}, this relation can be rewritten as

\[m > \frac{7.68\times10^{-27} \, \text{eV}}{A - 0.18}. \tag{9} \label{Eq. Mass Bounds}\]
The dependence on $A$ arises because the detectability of a periodic signal depends on its density contrast: higher-contrast periodicities are more readily detected and can therefore be constrained at smaller wavelengths (corresponding to larger particle masses), whereas lower-contrast periodicities become increasingly difficult to detect. The transformed relation presented here is valid over the range $0.2\,\text{Mpc}\lesssim \lambda_0 \lesssim 2\,\text{Mpc}$, corresponding to $0.212 \lesssim A \lesssim 0.5$.

We emphasise that the primary contribution of this work is the development and validation of a methodology for detecting FDM-induced periodicities, rather than the placement of competitive constrains on the FDM particle mass. The constraints derived in this appendix are necessarily conditional on the assumed correspondence between the measured periodicity and the de Broglie wavelength, and establishing the precise relationship between these quantities will require future simulation work. The methodology presented here provides a framework for translating measurements of structural periodicities into constraints on FDM particle masses once such a relationship is better understood. The sensitivity of this approach is ultimately determined by the smallest wavelength that can be robustly resolved observationally, which in turn depends primarily on the galaxy number density. As surveys such as DESI and 4MOST supersede SDSS, their increased sampling densities will enable progressively smaller wavelengths to be probed, extending the reach of this methodology to higher FDM masses.


\bsp	
\label{lastpage}
\end{document}